# Light ellipticity and polarization angle dependence of magnetic resonances in rubidium vapor using amplitude-modulated light: Theoretical and experimental investigations

Raghwinder Singh Grewal, Gour Pati, Renu Tripathi[*]

*Division of Physics, Engineering, Mathematics and Computer Science*
*Delaware State University, Dover, DE 19901, USA*

We report on experimental and theoretical investigations of the polarization dependence of magnetic resonance generated by synchronous optical pumping. Magnetic resonances with narrow linewidth are generated experimentally using a rubidium vapor cell with octade-cyltrichlorosilane (OTS) antirelaxation coating on inner walls. We studied the effect of light ellipticity on the amplitudes and widths of magnetic resonances by matching the light modulation frequency with $2\Omega_L$ (alignment) and $\Omega_L$ (orientation) in a Bell-Bloom interaction geometry, where $\Omega_L$ corresponds to the Larmor frequency. Both $2\Omega_L$ and $\Omega_L$ resonance amplitudes showed a strong dependence on the light ellipticity. In addition, we showed that the duty cycle of light modulation changes the slope of amplitude variations in $2\Omega_L$ and $\Omega_L$ resonances with light ellipticity. As a potential application, we showed that the difference between $2\Omega_L$ and $\Omega_L$ resonance amplitudes can be used for *in situ* measurement of light ellipticity. We also studied the dependence of $2\Omega_L$ and $\Omega_L$ resonance amplitudes on the polarization angle of linearly polarized light. These amplitudes oscillate periodically with the polarization angle. We found this oscillatory behavior to be sensitive to the tilt in magnetic field direction from the polarization plane. Such a property could be used to realize a vector magnetometer. A density matrix based theoretical model is developed to simulate the magnetic resonance spectrum for different light polarizations. Our theoretical model accurately reproduces the above mentioned experimental observations.

## I. INTRODUCTION

Coherent excitation of atomic ensemble with resonant light produces atomic coherence (or atomic polarization) in the ground-state of the medium. This causes interesting nonlinear magneto-optical phenomena such as Hanle effect [1–3] and nonlinear magneto-optical rotation (NMOR) [4,5] in the presence of magnetic field. Typically, these phenomena are observed by monitoring light transmission [6], scattered-light intensity [7] or light polarization angle [8] around the zero-magnetic field. Synchronous optical pumping is a well-established technique in which atomic coherence is produced by optically pumping of atoms at Larmor frequency, to create magnetic resonance at nonzero magnetic field. It is implemented by modulating the light amplitude [9–11], frequency [12,13] or polarization [14–16]. This method has been widely used in atomic magnetometry [17,18] and also, in precision measurement of permanent electric-dipole moment [19–21]

In synchronous optical pumping experiments, a spectrum containing multiple magnetic resonances is produced due to interaction of atoms with the frequency sidebands of modulated light [9,11]. Light polarization decides the type of magnetic resonance produced in the spectrum. Primarily, two types of magnetic resonances are produced due to (i) precession of atomic dipole moment (known as 'orientation') and (ii) precession of atomic quadrupole moment (known as 'alignment') [22–24]. In NMOR, linearly polarized light is used to produce 'alignment' resonance in a Faraday geometry where the magnetic field is applied along the light propagation direction [12]. Balanced polarimetric detection is used in NMOR to measure polarization angle with high sensitivity [17,18,25]. Both frequency and amplitude modulation of light have been explored in NMOR magnetometry. Most synchronous optical pumping experiments to date have been performed using either circular or linearly polarized light, after it was first demonstrated by Bell and Bloom [26] using amplitude modulated light and by applying magnetic field perpendicular to the light propagation direction. Unlike NMOR, Bell-Bloom technique produces an 'orientation' resonance using circularly polarized light. Recently, this technique has been extended for remote detection of geomagnetic field by exciting sodium atoms in the mesosphere [26–29].

It is known that the sensitivity of an optical magnetometer depends on the relative orientation between light propagation direction and magnetic field direction. For example, using Bell-Bloom technique, a longitudinal magnetic field parallel to the light propagation direction cannot produce a magnetic resonance (considered a 'dead-zone'). Studies have shown that periodic modulation of light polarization between two orthogonal states (right- and left-circular or orthogonal linear) can eliminate the dead-zone problem by simultaneously producing 'alignment' and 'orientation' resonances [16,30]. Alternatively, excitation using elliptical light polarization could improve the directional response of the optical magnetometer [31]. However, the effect of light ellipticity on magnetic resonances produced by synchronous optical pumping has not been very well studied.

[*]rtripathi@desu.edu

In this work, we present our studies on the dependence of magnetic resonance spectrum on light ellipticity and polarization angle. A rubidium (Rb) vapor cell with octade-cyltrichlorosilane (OTS) antirelaxation coating has been used in our experiment to produce narrow linewidth magnetic resonances. We show that excitation using elliptically polarized light can simultaneously produce $2\Omega_L$ (alignment) and $\Omega_L$ (orientation) resonances in a Bell-Bloom interaction geometry. We have developed a theoretical model to calculate the resonance spectrum using light with varying degree of ellipticity. Amplitudes of $2\Omega_L$ and $\Omega_L$ resonances show a strong dependence on the light ellipticity. Further, our study shows that the duty cycle of light modulation can effectively control the amplitude variations of $2\Omega_L$ and $\Omega_L$ resonances with ellipticity. Finally, we have presented a simple scheme for vector magnetometry to determine magnetic field direction by measuring the oscillations in $2\Omega_L$ and $\Omega_L$ resonance amplitudes with light polarization angle.

This paper is organized as follows. In Section II below, we discuss the theoretical model developed for studying the dependence of magnetic resonance spectrum on the light ellipticity and the polarization angle. A description of our experimental setup is provided in Sec. III. Results and discussions including comparisons between experimental and theoretical results are presented in Sec. IV.

## II. THEORETICAL MODEL

Our theoretical model is based on atomic density matrix equations. We consider a resonant excitation between the hyperfine ground state $F_g = 3$ and the hyperfine excited state $F_e = 2$ in $^{85}$Rb $D_1$ manifold, including all Zeeman sublevels in each hyperfine state. To study magnetic resonances, we considered a laser field propagating along the z-axis:

$$\vec{E}(t) = \hat{e}_L E_o e^{-i\omega t} + c.c. \tag{1}$$

where $E_o$ is the amplitude of laser field with resonant frequency $\omega$. The polarization vector $\hat{e}_L$ of the field is described by

$$\hat{e}_L = \hat{e}_x \cos\varepsilon + i\hat{e}_y \sin\varepsilon \tag{2}$$

where $\hat{e}_x$ and $\hat{e}_y$ are the unit vectors along x- and y-axes, respectively and $\varepsilon$ is the ellipticity angle of the laser field. All polarization states of the laser can be described by a value of $\varepsilon$ ranging between 0º to 45º. For our modeling purposes, we have considered (i) linear ($\varepsilon = 0°$) and (ii) elliptical polarizations ($\varepsilon \neq 0°$) of the optical field corresponding to a transition from ground state $F_g = 3$ to excited state $F_e = 2$. Fig. 1 depicts both of these scenarios. In Fig. 1(a) (inset), for linearly polarized light, we consider the light propagation along $\hat{z}$ direction, electric field vector is chosen along the y-axis and a magnetic field B is considered parallel to the x-axis ($B = B_x$). To give a physical picture of the light-atom interaction, we choose the axis of quantization to be along the magnetic field direction (i.e. x-axis). In the presence of magnetic field, the Zeeman sublevels shift by integer multiples of the Larmor frequency $\Omega_L$ i.e. $\pm m_F \Omega_L (m_F \gamma_{Rb} B_x)$, where $\gamma_{Rb}$ is the Gyromagnetic ratio of Rb atom and $m_F$ is the magnetic quantum number of the sublevels. Since the linearly polarized field is chosen perpendicular to the quantization axis, it produces equal $\sigma^+$ and $\sigma^-$ transitions corresponding to $\Delta m_{eg} = \pm 1$ between the ground and excited state sublevels, as shown in Fig. 1(a).

These two transitions create a three-level Λ-system involving one common excited state and two ground state Zeeman sublevels with $\Delta m_g = 2$, shown using a curved arrow in Fig. 1(a). A magnetic resonance is formed by coherent population trapping (CPT) due to a dark superposition of the participating ground state Zeeman sublevels. Magnetic resonance can also be produced by other Λ-systems (not shown in the figure) that could be formed in $F_g$ =3 →$F_e$=2 transition involving other Zeeman sublevels of $F_g$ and $F_e$ that satisfy the same condition $|\Delta m_g| = 2$. The transitions corresponding to elliptically polarized light are shown in Fig. 1(b). In this case, an additional π-transition corresponding to $\Delta m_{eg} = 0$ is introduced between the ground and excited state sublevels when the light electric field vector becomes parallel to the quantization axis. This creates additional Λ-systems involving a superposition of ground state sublevels satisfying $\Delta m_g = \pm 1$ [shown with curved arrows in Fig. 1(b)]. The strength of a $\pi$-transition increases with light ellipticity and becomes equal to the circular $\sigma$ components at $\varepsilon = 45°$.

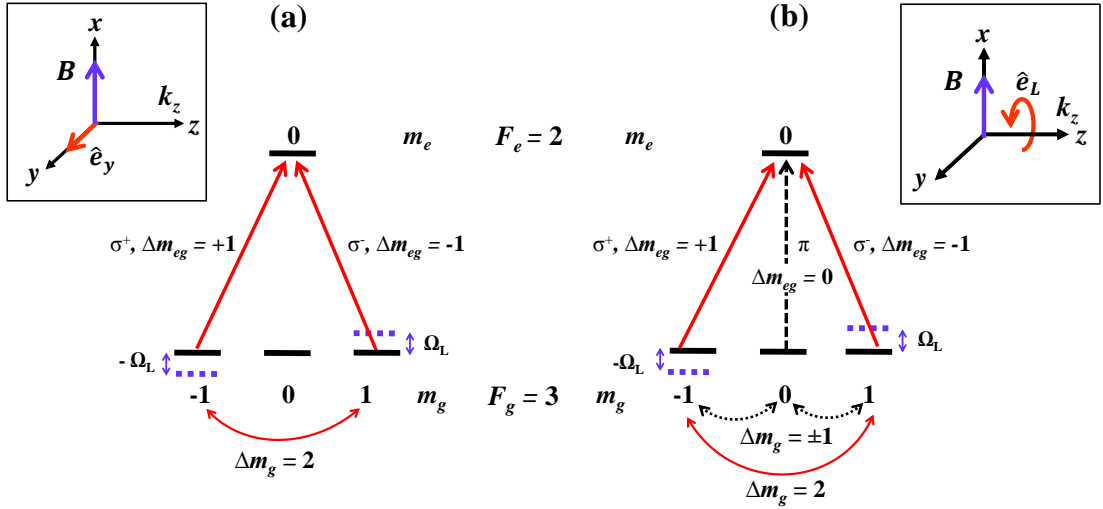

FIG. 1. A two-level atomic system $F_g=3 \to F_e=2$ interacting with (a) linear and (b) elliptical polarized light. Here 'g' refers to the ground, and 'e' to the excited state. $\Omega_L$ is the Larmor frequency. $m_g$ and $m_e$ are the magnetic quantum number of the ground and excited state sublevels, respectively. $\Delta m_g$ is the coherence among the ground state sublevels and $\Delta m_{eg}$ represent the coherence between ground and excited state sublevels. The axis of quantization is considered parallel to the magnetic field $\boldsymbol{B}$ (i.e. along the $x$-axis) as shown in the insets (a) and (b).

In the synchronized optical pumping scheme, the amplitude of the laser field is modulated using a square-wave modulation function $\xi(t)$ with a duty cycle $\eta$. The laser field with amplitude modulation (AM) in the model is described as

$$\vec{E}(t) = \hat{e}_L E_o (\eta + \Delta M \, \xi(t)) e^{-i\omega t} + c.c. \tag{3}$$

where $\Delta M$ is the modulation depth. The Fourier series expansion of the square-wave function $\xi(t)$ can be written as

$$\xi(t) = \sum_{m=1}^{m=\infty} g_m(\eta) \cos(m\Omega_{mod} t) \tag{4}$$

$$g_m(\eta) = \frac{2}{m\pi} \sin[m\pi\eta] \tag{5}$$

Here, the terms $g_m$ are Fourier coefficients of the square-wave modulation function $\xi(t)$ and the integer index $m$ corresponds to different harmonics of $\xi(t)$. At a fixed modulation frequency $\Omega_{mod}$, the optical field consists of frequency sidebands at $\omega \pm m\Omega_{mod}$ along with a carrier at the laser frequency $\omega$. The amplitude $g_m$ of a particular sideband depends on duty cycle $\eta$ of $\xi(t)$ [32]. For example, at 50% duty cycle ($\eta = 0.5$), $g_m$ value for all even integer harmonics (even values of $m$) will be equal to zero and odd harmonics (odd values of $m$) will be nonzero. The modulated field has dominant first-order sidebands at frequencies $\omega \pm \Omega_{mod}$ for which $g_m$ value is maximum. These sidebands in the modulated field cause synchronous pumping of the atoms to create multiple magnetic resonances at nonzero magnetic field satisfying the condition

$$\Omega_{mod} = k\Omega_L \tag{6}$$

where $k$ is the rank of the atomic spin polarization moment of the density matrix. These moments of the angular momentum state $F_g$ are related to coupling (or coherence) between Zeeman sublevels with $\Delta m_g (\neq 0)$ [23,33]. For a given choice of the quantization axis, these coherences contribute to all atomic polarization moments with $k \geq |\Delta m_g|$ and have a maximum possible rank $k$ equal to $|\Delta m_g| = 2F_g$. Generation and detection of multipole moments of rank $k \leq 2$ could be accomplished by using weak light, whereas higher rank moments ($k > 2$) require multiphoton interactions [33]. Our study is focused on magnetic resonances formed due to multipole moments of rank $k \leq 2$ at different light ellipticities. For linearly polarized light [Fig.1 (a)], the coherence condition $|\Delta m_g| = k = 2$ (quadrupole moment, also known as 'alignment') creates a resonance at $\Omega_L = \Omega_{mod}/2$. For elliptically polarized light [Fig.1 (b)], the coherence conditions $|\Delta m_g| = k = 2$ (quadrupole moment) and $|\Delta m_g| = k = 1$ (dipole moment, also known as 'orientation') create two resonances at $\Omega_L = \Omega_{mod}/2$ and $\Omega_L = \Omega_{mod}$, respectively. For an arbitrary duty cycle $\eta$ ($\neq 0.5$), each frequency sideband corresponding to integer index $m$ produces two resonances at $\Omega_L = \pm m\Omega_{mod}/k$ corresponding to $k = 2$ and $k = 1$. As a general rule, when elliptically polarized light is used and magnetic field $B_x$ is varied, resonances may occur at following frequencies:

$$
\begin{array}{cccccc}
m & = & 1 & 2 & 3 & 4 & \ldots \\
\Omega_L^{k=2} & = & \pm\dfrac{\Omega_{mod}}{2} & \pm\Omega_{mod} & \pm\dfrac{3\Omega_{mod}}{2} & \pm 2\Omega_{mod} \ldots \\
\\
\Omega_L^{k=1} & = & \pm\Omega_{mod} & \pm 2\Omega_{mod} & \pm 3\Omega_{mod} & \pm 4\Omega_{mod} \ldots
\end{array}
\tag{7}
$$

The resonance condition for a particular harmonic for $k = 1$ could also match with the one for a different harmonic for $k = 2$. For example, resonances for $k = 2, m = 2$ and $k = 1, m = 1$ occur simultaneously at $\Omega_L = \pm\Omega_{mod}$. The combined resonances due to both $k = 1$ and 2 can be written as

$$
\begin{array}{ccccccc}
\Omega_L = & \pm\dfrac{\Omega_{mod}}{2}, & \pm\Omega_{mod}, & \pm\dfrac{3\Omega_{mod}}{2}, & \pm 2\Omega_{mod}, & \pm\dfrac{5\Omega_{mod}}{2}, & \pm 3\Omega_{mod} \ldots \\
n & \pm 1 & \pm 2 & \pm 3 & \pm 4 & \pm 5 & \pm 6
\end{array}
\tag{8}
$$

Here, the integer index $n$ is used to label the position of all resonances formed by elliptically polarized light modulated with an arbitrary duty cycle $\eta$. The resonance at zero magnetic field (Hanle resonance) due to the carrier is represented by $n = 0$.

We calculate the magnetic resonances using the time evolution of atomic density matrix $\rho$ given by the Liouville equation:

$$\frac{\partial \rho}{\partial t} = \frac{1}{i\hbar}\left[\tilde{H}, \rho\right] - \frac{1}{2}\left\{\hat{\Gamma}, \rho\right\} + R \tag{9}$$

Here $\tilde{H}$ represents the total Hamiltonian of the atomic system in the rotating wave frame. It includes the internal atomic energy levels, light-atom interaction, and magnetic field-atom interaction. The diagonal relaxation matrix $\hat{\Gamma}$ includes the spontaneous decay rate $\Gamma$ of the excited state and the transit relaxation rate $\gamma$ of each sublevel due to exit of atoms from the laser beam. Matrix $R$ describes the repopulation of ground state sublevels due to decay rates

$\Gamma$ and $\gamma$. The theoretical model is simplified by not considering the atomic motion (or velocity distribution), the effect of neighboring transitions, and the spatial distribution of laser intensity. Time-dependent density matrix equations obtained from the Eq. (9) are solved numerically to calculate light absorption coefficient of the medium for ellipticity $\varepsilon$ using the following expression:

$$\alpha(t) \approx \sum_{e_i g_j} \frac{\Gamma}{\pi \Omega_R} \left( \beta_{e_i g_j} \cos[\varepsilon] \, \text{Im}[\rho_{e_i g_j}(t)] + \beta'_{e_i g_j} \sin[\varepsilon] \, \text{Re}[\rho_{e_i g_j}(t)] \right) \qquad (10)$$

where $\Omega_R = \langle F_g \| D \| F_e \rangle E_o$ is the reduced Rabi frequency of the laser field and $D$ is the dipole operator. The ground state and excited state sublevels involved in the optical transition decide the strength of coefficients $\beta_{e_i g_j}$ and $\beta'_{e_i g_j}$. To keep consistency with our experimental observation, we calculate the power $P(t)$ in transmission using the following expression:

$$P(t) = P_0 e^{-\alpha(t)L} \approx P_0 (1 - \alpha(t)L) \qquad (11)$$

Here, $P_0$ is the initial laser power which is set to unity to simplify our calculations. We have also assumed that $\alpha L \ll 1$ which is true for an optically thin medium. The in-phase and quadrature components of the magnetic resonances are calculated by demodulating $\alpha(t)$ [eq. (10)] and $P(t)$ [eq. (11)] at the first-harmonic of $\pm\Omega_{mod}$ [34].

### III. EXPERIMENTAL SETUP

A schematic diagram of the experimental setup is shown in Figure 2. A tunable external cavity diode laser with resonant wavelength 795 nm matching the $^{85}$Rb $D_1$ transition and linewidth less than 500 kHz is used in the experiment. Using a combination of half-wave plate $\lambda/2$ and a polarizing beam splitter (PBS), the laser beam is split into two paths. The reflected beam from the PBS is passed through an acoustic-optic modulator (AOM1) driven by 80 MHz radio-frequency signal of fixed amplitude. The first-order diffracted beam from the AOM1 is utilized in a saturation absorption spectroscopy (SAS) setup [35]. The laser is locked to $F_g= 3 \rightarrow F_e= 2$ transition of the $^{85}$Rb $D_1$ line using the Doppler-free absorption peak produced in the SAS setup. Light transmitted through PBS is amplitude-modulated via AOM2 driven by 80 MHz radio-frequency signal using a rectangular pulse waveform generated through a function generator with an arbitrary duty cycle $\eta$. The diameter of first-order diffracted beam is expanded from 2 mm to 8 mm using a couple of lenses in a telescopic configuration. The expanded beam increases the interaction time of atoms with the laser beam. Laser power to the vapor cell is controlled using a neutral density (ND) filter. Light ellipticity is varied from 0º to 45º using a λ/4 plate placed in the beam path. A buffer gas free OTS-coated rubidium vapor cell (length = 2 cm, diameter = 2.5 cm) obtained from Precision Glassblowing is mounted at the center of a four-layer magnetic shield (MS) with a shielding factor of ~$10^6$. The MS contains a printed three-axis magnetic field coil installed inside its innermost layer. The coils are connected to three independent current sources to independently apply static and/or scanning magnetic field in any arbitrary direction as required in the experiment. The Rb vapor cell is kept at room temperature. Light transmitted through the cell is detected using a photodiode (PD). The OTS coating in the Rb vapor cell allows us to produce narrow linewidth magnetic resonances by reducing the effect of wall collisions and thereby reducing the depolarization of Rb atoms [36,37]. Laser excitation of $F_g = 3 \rightarrow F_e = 2$ transition creates magnetic resonances with approximately 3% contrast, which are measured by demodulating the PD output using a lock-in amplifier operating at the first harmonic of the laser modulation frequency $\Omega_{mod}$. The two channels of the lock-in amplifier allow us to measure simultaneously the in-phase (or amplitude) and quadrature (or phase) components of the magnetic resonances.

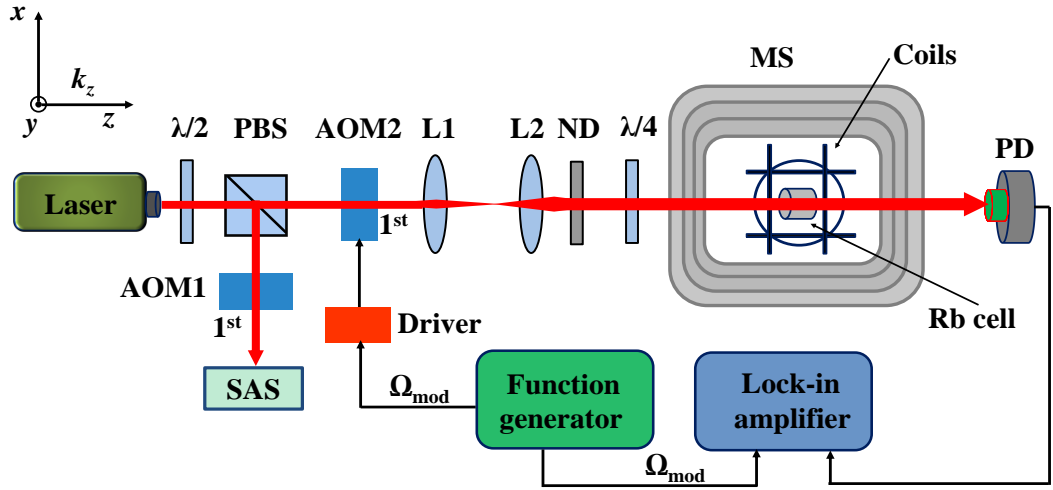

FIG. 2. Schematic diagram of the experimental setup. $\lambda/2$, half-wave plate; $\lambda/4$, quarter-wave plate; L1-L2, convex lenses; PBS, polarizer beam splitter; MS, magnetic shield; ND, neutral density filter; PD, photodiode; AOM1-2, acoustic optic modulator; SAS, saturation absorption spectroscopy. The choice of our coordinate system is shown in the figure indent.

## IV. RESULTS AND DISCUSSION

### A. Dependence of magnetic resonances on the light ellipticity

Figure 3 (a) shows the in-phase and quadrature components of experimentally observed magnetic resonances in the transmitted light for three different values (0º, 15º and 45º) of the light ellipticity. These measurements are performed by scanning the transverse magnetic field $B_x$ around the zero-field. Modulation frequency of the laser is kept fixed at $\Omega_{mod} = 3$ kHz with duty cycle $\eta = 0.5$. Average intensity of the laser beam is set to 0.2 mW/cm². For zero ellipticity ($\varepsilon = 0º$) corresponding to linearly polarized light, the in-phase signal shows a resonance (labeled as $n = 0$) around the zero magnetic field. As explained in section II, the $n=0$ resonance occurs due to coupling between degenerate ground state Zeeman sublevels with $|\Delta m_g|=2$ produced by the $\sigma^+$ and $\sigma^-$ polarization components of the carrier at laser frequency $\omega$. The presence of a peak suggests a dark resonance, as expected for $F_g \rightarrow F_e = F_g -1$ transition [3]. The in-phase and quadrature signals also show resonances at $\Omega_L = \pm \Omega_{mod}/2$, the so-called $n=\pm 1$ resonances, due to first-order sidebands at $\omega \pm \Omega_{mod}$ of the modulated laser field. These two sidebands along with the carrier introduce coupling between ground state Zeeman sublevels with $|\Delta m_g|=2$ to produce $n=\pm 1$ resonances at non-zero magnetic field. Alternatively, $n=\pm 1$ 'alignment' resonances can be described as being produced by synchronous pumping of atoms with modulated light at Larmor frequency with atomic polarization moment $k=2$. Since the magnetic field direction is transverse to the light propagation direction, resonances labeled as $n=0$ and $\pm 1$ cannot be observed in polarization rotation measured using a balanced polarimeter setup, i.e. the amplitude of polarization rotation decreases as a cosine of angle between the light propagation direction and the magnetic field [38].

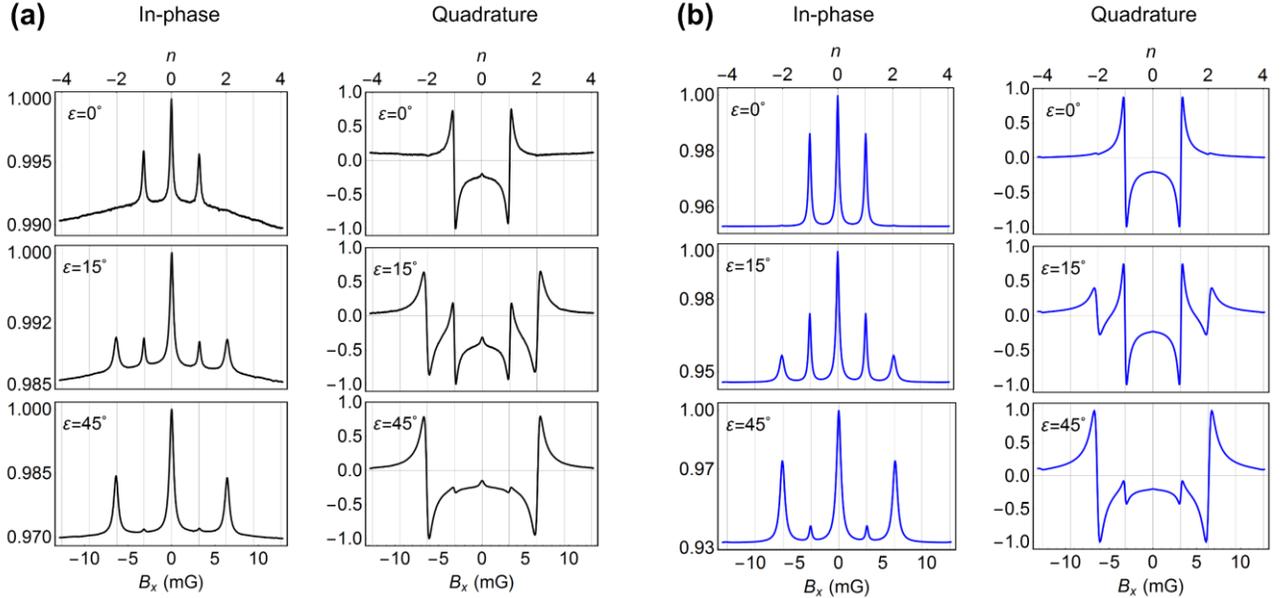

FIG. 3. Experimentally measured (a) and theoretically calculated (b) magnetic resonances using modulated light ($\eta = 0.5$ and $\Omega_{mod} = 3$ kHz) with different light ellipticity values in each row. Average laser intensity in the experiment is set to 0.2 mW/cm². Parameters used in simulations: $\Omega_R = 0.01\ \Gamma$, $\gamma = 3 \times 10^{-5}\ \Gamma$. Plots are normalized with respect to the maximum amplitude in the respective signal. All the resonances are labelled using the integer index $n$ described in section II.

As light ellipticity is increased from zero, the $\pi$ and $\sigma^\pm$ components tend to form additional $\Lambda$-type systems satisfying the condition $|\Delta m_g|=1$ between the ground state Zeeman sublevels. Therefore, for non-zero ellipticity, both $|\Delta m_g|=1$ and $|\Delta m_g|=2$ coherence conditions will contribute to produce resonances at zero magnetic field. Due to the presence of sidebands, coherence condition $|\Delta m_g|=1$ creates 'orientation' resonances at $\Omega_L = \pm \Omega_{mod}$ labeled as $n=\pm 2$ with polarization moment $k=1$ [Fig. 3 (a) (middle row)]. At light ellipticity $\varepsilon = 15^\circ$, the amplitudes of $n=\pm 1$ and $n=\pm 2$ resonances are approximately equal in both in-phase and quadrature signals. When ellipticity is further increased, the amplitudes of $n=\pm 1$ resonances decrease and nearly vanish at $\varepsilon = 45^\circ$ corresponding to circularly polarized light [Fig. 3 (a) (bottom row)]. On the other hand, the amplitudes of $n=\pm 2$ resonances increase with increase in ellipticity and become maximum at $\varepsilon = 45^\circ$. This behavior can be inferred from the increase in strength of the $\pi$-transition responsible for $|\Delta m_g|=1$ coherence with increase in ellipticity. Theoretical results shown in Fig. 3(b) show a good agreement with our experimental observations shown in Fig. 3 (a). However, we observed a difference between theory and experiment in terms of the relative amplitudes of resonances for each ellipticity case. For example, unlike the experimental results shown in Fig. 3 (a) (middle row), the calculated resonances $n=\pm 1$ and $n=\pm 2$ in Fig. 3 (b) (middle row) do not have equal amplitudes for ellipticity $\varepsilon = 15^\circ$. This discrepancy could have resulted from the simplification of our theoretical model discussed in section II.

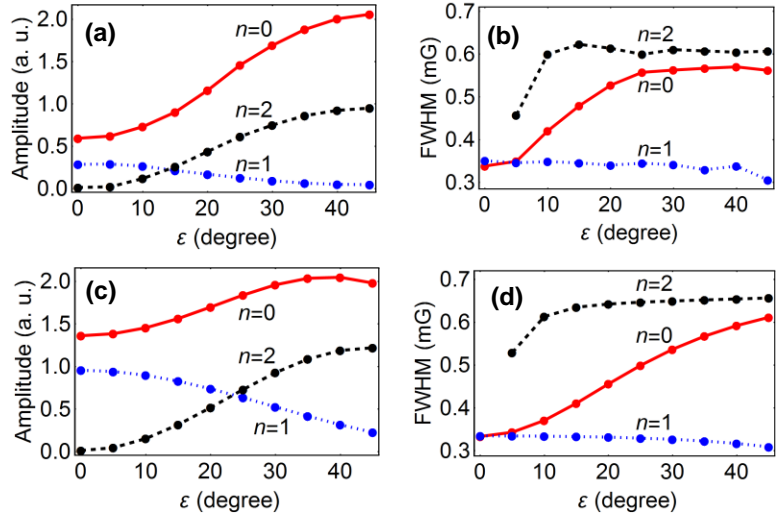

FIG. 4. Experimentally measured (first row) peak amplitudes (a) and FWHM widths (b) of magnetic resonances in the in-phase signal as a function of the light ellipticity. Corresponding theoretical results are shown in the second row (c) and (d). The experimental and theoretical parameters are similar to that chosen for the results presented in Fig. 3.

Figure 4(a) shows experimentally measured peak amplitudes of in-phase $n = 0$, 1 and 2 resonances as a function of light ellipticity $\varepsilon$ for duty cycle $\eta = 0.5$. Since zero-field $n = 0$ resonance has contributions from many degenerate ground state superpositions of sublevels satisfying conditions $|\Delta m_g| = 1$ and/or $|\Delta m_g| = 2$, it has much higher amplitude compared to $n = 1$ and $n = 2$ resonances. Increase in the amplitude of $n = 0$ resonance with ellipticity, is in agreement with previous reports where continuous laser excitation was utilized [39,40]. Since the strength of $|\Delta m_g| = 2$ coherence weakens with an increase in ellipticity from 0º to 45º, the amplitude of $n = 1$ resonance consequently diminishes, as shown in Fig. 4(a). Figure 4(b) shows the plots of full-width half maximum (FWHM) of in-phase $n = 0$, 1 and 2 resonances as a function of light ellipticity. The width of a resonance depends on the dephasing between the ground state sublevels, and on the Rabi frequencies (or matrix elements/coupling strengths) of the optical transitions involved in a particular $\Lambda$-system. Closer to zero ellipticity, $n = 0$ and $n = 1$ resonances have approximately equal FWHMs. The width of $n = 0$ resonance is increased by nearly 60% from 0.35 mG to 0.56 mG when $\varepsilon$ is changed from 5º to 25º and becomes approximately constant thereafter for higher ellipticities. On the other hand, the width of $n = 1$ resonance does not vary much with ellipticity. The $n = 2$ resonance shows a broader linewidth compared to $n = 0$ and $n = 1$ resonances for all ellipticity values. Theoretical results shown in Fig. 4(c) and Fig. 4(d) are consistent with the corresponding experimental results shown in Fig. 4(a) and Fig. 4(b), respectively. The amplitudes and widths of $n = 1$ and $n = 2$ resonances were also measured from the quadrature signals, and found to exhibit similar dependencies on light ellipticity, as the ones measured from in-phase signals.

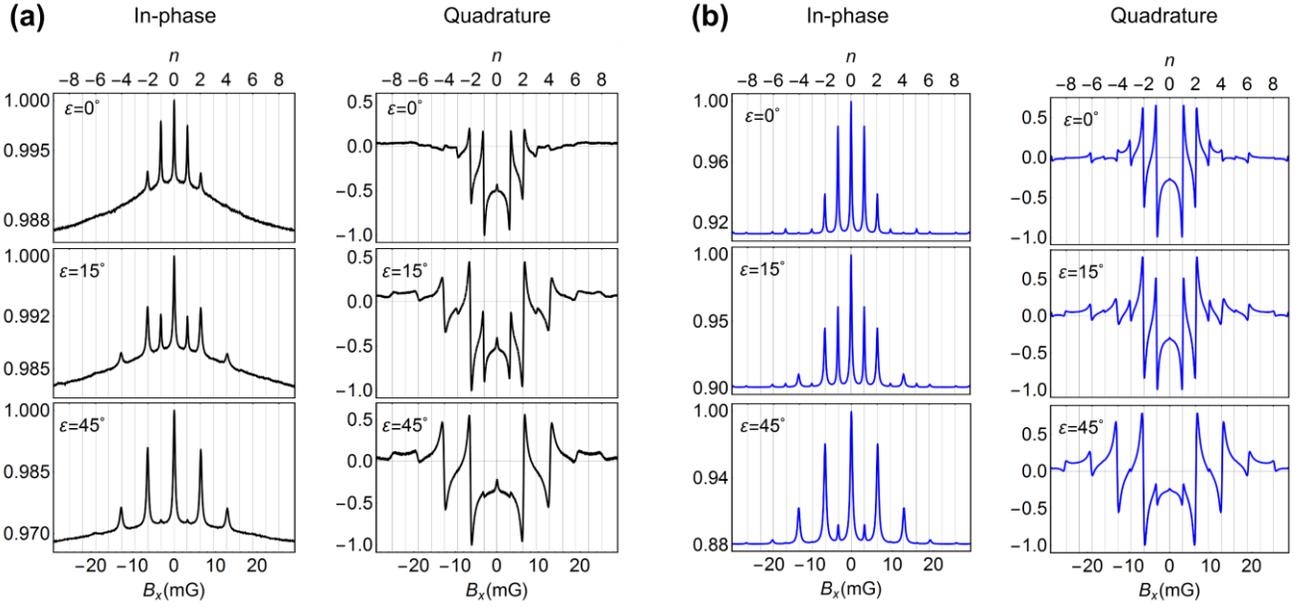

FIG. 5. Experimentally measured (a) and theoretical calculated (b) magnetic resonances for different light ellipticities (as labelled) using modulated light with $\eta = 0.3$ and $\Omega_{mod} = 3$ kHz. Other parameters used in simulations remain the same as in Fig. 3. Plots are normalized with respect to the maximum amplitude in respective signal. All the resonances are labelled using the integer index $n$ described in section II.

Next, we describe the effect of light ellipticity on the magnetic resonances for a lower duty cycle of light modulation $\eta = 0.3$ and by keeping the average laser intensity fixed at 0.2 mW/cm². In Fig. 5(a), higher-order magnetic resonances are observed due to the presence of all sidebands in the modulated light. For $\varepsilon = 0º$, the in-phase signal shows resonances up to second-order (i.e. $n = \pm 2$) and the quadrature signal which is phase-sensitive, shows resonances up to the fourth-order (i.e. $n = \pm 4$) satisfying the resonance condition of $\Omega_L = \pm m\Omega_{mod}/2$ with $m = 4$ (i.e. fourth sideband of modulated light) and $k = 2$. The quadrature signal shows a dispersive line shape with same sign for the first three-orders and an opposite sign for the fourth-order (i.e. $n = \pm 4$) resonances [Fig. 5(a) (top row)] indicating a 180° phase change, possibly due to a sign reversal of the Fourier coefficient $g_4$.

As the light ellipticity is changed from zero, resonances up to eighth-order (i.e. $n = \pm 8$) are observed [Fig. 5(a)]. As discussed in Section II, the $k = 1$ resonance (for which $\Omega_L = m\Omega_{mod}$) and the $k = 2$ resonance (for which $\Omega_L = m\Omega_{mod}/2$) are produced simultaneously due to nonzero light ellipticity. The $\Omega_L$ values for these $k$'s are satisfied by two different $m$ values that correspond to two different sidebands of the modulated light. The dominance of a particular side-band to form this type of higher-order resonance is decided by the strength of ground state coherence $|\Delta m_g|$. For example, $n = 4$ resonance at $\Omega_L = 2\Omega_{mod}$ is produced by the fourth sideband ($m = 4$) with polarization moment $k = |\Delta m_g|=2$ and also, by the second sideband ($m =2$) with polarization moment $k = |\Delta m_g|=1$. For a non-zero light ellipticity [Fig. 5a (middle and bottom row)], $n = 4$ dispersive resonance in the quadrature signal switches its sign with respect to $n = 4$ dispersive resonance for the zero light ellipticity case [Fig. 5(a) (top row)]. The switching of sign in $n = 4$ dispersive resonance with ellipticity is due to the dominance of the participating sideband from $m = 4$ to $m = 2$ (no sign reversal in $g_2$) satisfying the coherence condition $|\Delta m_g|=1$ for $k = 1$. Similarly, the inter-sign reversal of $n = 8$ dispersive resonance with respect to $n = 4$ dispersive resonance at ellipticity $\varepsilon = 15°$ or 45° can be explained by the fact that it is predominantly formed by the strong fourth sideband (i.e. $m = 4$ with a sign reversal in $g_4$) of the modulated light. Fig. 5(b) shows corresponding theoretical results obtained for the same light ellipticities as in Fig. 5(a). The theoretical results reproduce most of the salient features observed experimentally in Fig. 5(a). The sign-change of $n = 4$ and 8 resonances in quadrature signals discussed above, is also clearly observed in the simulated results shown in Fig. 5(b).

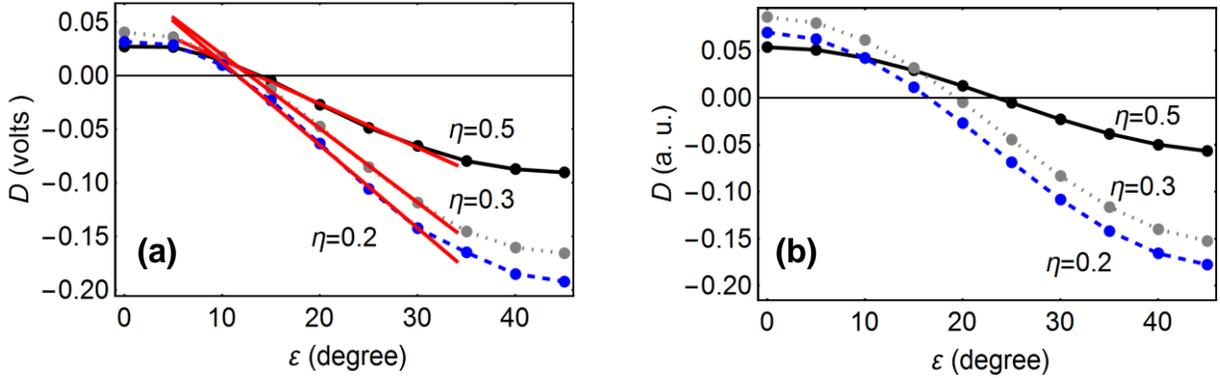

FIG. 6. Variation of experimentally measured (a) and theoretically calculated (b) difference $D$ in amplitudes of $n =1$ and $n = 2$ resonances as a function of light ellipticity for different values of $\eta$. Other parameters used in the simulation are same as those described in Fig. 3.

Next, we measured the peak amplitudes of in-phase $n =1$ and $n = 2$ resonances as a function of the light ellipticity for arbitrary duty cycle $\eta$ of the modulated light. The variation in amplitudes of these resonances with ellipticity is found to change with duty cycle $\eta$. To study this dependence, we measured the amplitude difference ($D$) of $n =1$ and $n = 2$ resonances as a function of the light ellipticity for different values of $\eta$. This is shown in Fig. 6(a). For $\eta = 0.5$, the $D$ value varies from positive to negative and passes through a zero-crossing around $\varepsilon = 15°$ where $n =1$ and $n = 2$ resonance amplitudes are equal. This is consistent with our results shown in Fig. 4(a). The zero-crossing shifts towards a lower ellipticity [ i.e. $\varepsilon = 11.7°$] with a decrease in $\eta$ from 0.5 to 0.3 [Fig. 6(a)]. The plots in Fig. 6(a) show that the difference $D$ varies quite linearly over ellipticity $\varepsilon$ ranging from 10° to 30° at three different duty cycles $\eta$. The red lines show linear fittings to the experimental data from which the slope $\partial D/\partial \varepsilon$ is calculated. The slope $|\partial D/\partial \varepsilon|$ increases from 4.1 mV/deg to 7.7 mV/deg by changing $\eta$ from 0.5 to 0.3. A curve with highest slope can be utilized in applications that require *in situ* measurement of the light ellipticity with higher accuracy. Precise measurement of light ellipticity is crucial for many atom-based systems. For example, level shifts controlled by light polarization in an optical lattice can be used to implement quantum logic gates, and can utilized to estimate frequency error or accuracy of an optical lattice clock [41,42]. A device for light ellipticity measurement can be realized by doing numerical data fitting to find peak amplitudes of the resonances and their difference $D$, and extracting ellipticity information from a calibration curve shown in Fig. 6a. Figure 6(b) gives theoretical plots showing variations in $D$ with ellipticity for the same $\eta$ values. Theoretical results show a good agreement with the experiment, particularly, in reflecting the increase in slope $|\partial D/\partial \varepsilon|$ with lowering of duty cycle $\eta$. However, the zero-crossing points for theoretical plots in Fig. 6(b) do not match with those in Fig. 6(a) due to the simplification of our theoretical model.

### B. Polarization angle dependence of magnetic resonances and determination of magnetic field direction

So far, we have discussed the effect of light ellipticity on the non-zero magnetic resonances formed by σ and π transitions in the presence of a transverse magnetic field along the *x*-axis. The σ and π transitions can also be produced by changing the relative angle between the magnetic field (i.e. direction of axis of quantization) and polarization vector of a linearly polarized light. Next, we discuss the amplitude dependence of these resonances on the direction of the magnetic field. For this study, we kept the magnitude of total magnetic field fixed at $B = 5.1$ mG (i.e. $\Omega_L = 2.4$ kHz) and measured amplitudes of $n =1$ and $n = 2$ resonances with polarization angle by changing the magnetic field direction. In this case, magnetic resonances are observed by scanning the modulation frequency $\Omega_{mod}$ of light from 1.5 kHz to 6 kHz. Average laser intensity is kept fixed at 0.2 mW/cm$^2$ and the duty cycle $\eta$ of light modulation is set at 50% (i.e. $\eta = 0.5$). Linearly polarized light is used in this study. The orientation of the polarization vector $E$ with respect to the magnetic field $B$ is varied to induce coherence $|\Delta m_g|$ for producing $n =1$ and $n = 2$ resonances.

Figure 7(a) shows the geometrical representation of the polarization vector $E$ and the magnetic field $B$ in a three-dimensional coordinate system. As shown, the polarization vector makes an angle $\phi$ with the *y*-axis in the *x-y* plane. The direction of the magnetic field $B$ is defined by angles $\theta$ and $\psi$ in the spherical coordinate system, where $\psi$ is the azimuthal angle and $\theta$ is the polar angle. To illustrate the concept, we set the azimuthal angle of the magnetic field $\psi = 30°$ for all our measurements. For a given angle $\theta$ of the $B$ field, the peak amplitudes of resonances at $\Omega_{mod} = \Omega_L = 2.4$ kHz (i.e. $n = 2$) and $\Omega_{mod} = 2\Omega_L = 4.8$ kHz (i.e. $n=1$) are measured by changing the rotation angle $\phi$ of the polarization vector $E$. The amplitudes of $2\Omega_L$ and $\Omega_L$ resonances depend on the strength of σ- and π- transitions, which are determined by an angle between polarization vector and orientation of the magnetic field (i.e. axis of quantization), as discussed in section II.

Figure 7(b) shows the amplitudes of $2\Omega_L$ and $\Omega_L$ resonances as a function of the angle $\phi$ corresponding to three different values of $\theta$. For $\theta = 90°$, the magnetic field $B$ is oriented in the *x-y* plane i.e. in the plane of light polarization [Fig. 7(a)]. When the polarization angle $\phi = 30°$, the electric field vector $E$ becomes parallel to $B$ resulting in only a π -transition between the Zeeman sublevels of ground and excited states. In this case, due to the absence of σ transition, no coupling (or coherence) can be established between the ground state sublevels. Thus, amplitudes of both $2\Omega_L$ and $\Omega_L$ resonances become zero at the polarization angle $\phi = 30°$. This can be seen in the plots shown in the top row of Fig. 7(b). The amplitude of $2\Omega_L$ resonance shows a plateau for polarization angle $\phi$ between 10° and 60°. This is because it is only formed by σ-transitions which remain weak over this range of angle. On the other hand, the amplitude of $\Omega_L$ resonance (which is formed by both σ- and π-transitions) changes rapidly and reaches its maximum value at angle $\phi=75°$, where the strengths of σ and π components become equal as the angle difference $(\phi - \psi) = 45°$. When the polarization angle $\phi$ is changed further beyond 75°, the angle difference $(\phi - \psi)$ is increased beyond 45° resulting in a stronger σ-transition. At $\phi =120°$ where $(\phi - \psi) = 90°$, the light will only have σ components, which results in a maximum amplitude of $2\Omega_L$ resonance and minimum amplitude of $\Omega_L$ resonance [Fig. 7(b) (top row)]. The amplitude of $2\Omega_L$ resonance shows the next plateau around $\phi =210°$ due to a resulting π only transition. Thus, the amplitude of $2\Omega_L$ resonance oscillates slowly with a periodicity of 180° in the polarization angle $\phi$. On the other hand, the amplitude of $\Omega_L$ resonance oscillates faster with a periodicity of 90° in angle $\phi$ for the case $\theta = 90°$.

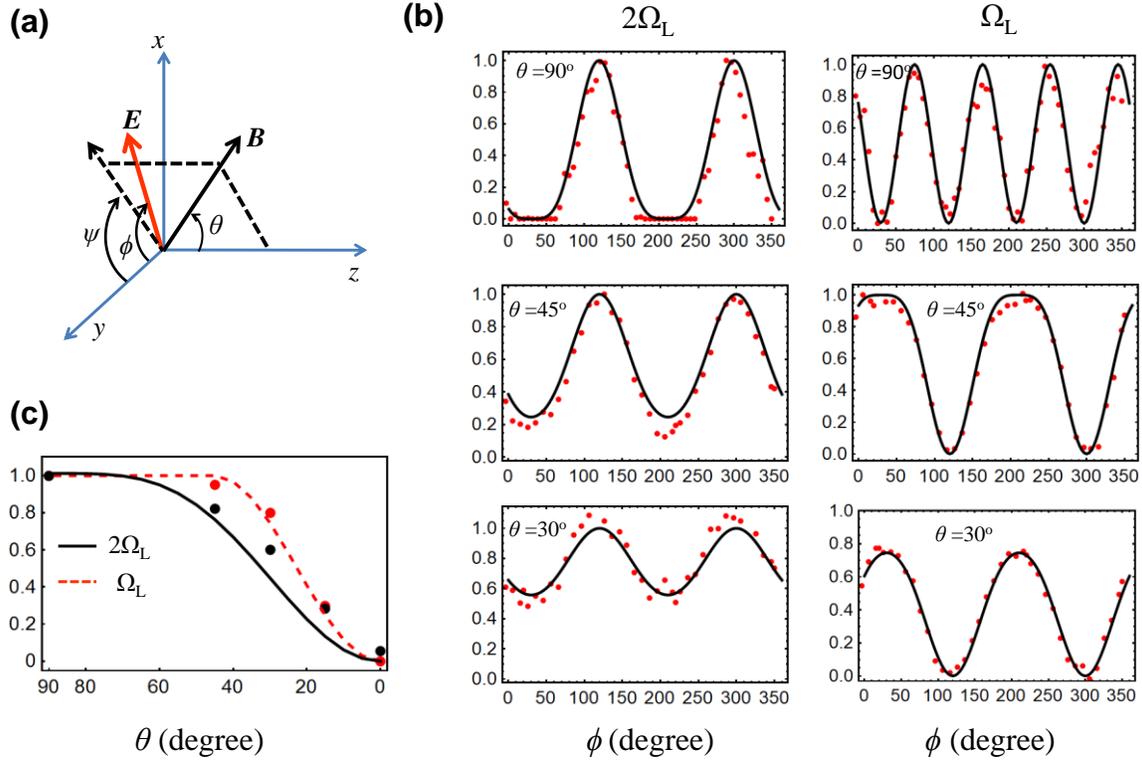

FIG. 7. (a) Geometrical representation of polarization vector $E$ and magnetic field vector $B$ in spherical coordinate system. (b) Amplitudes of $2\Omega_L$ and $\Omega_L$ resonances as a function of polarization rotation angle $\phi$ at different tilt angle $\theta$ of the magnetic field $B$. (c) Variations in oscillation amplitudes of $2\Omega_L$ and $\Omega_L$ resonances with $\theta$. Average laser intensity is fixed at 0.2 mW/cm$^2$ and modulation duty cycle $\eta = 0.5$. Experimentally measured (dots) and theoretically calculated (solid lines) data are normalized in all plots with respect to peak amplitude in $\theta = 90°$ (no tilt in $B$) case.

Next, we considered a tilt in the magnetic field direction (keeping its strength fixed) from the $x$-$y$ plane by defining the angle $\theta$ to be less than 90° [Fig. 7(a)]. In this case, rotation of polarization angle $\phi$ cannot make the electric field vector $E$ parallel to the magnetic field $B$ to produce only a $\pi$ transition. Therefore, unlike Fig. 7(b) (top row), polarization angles $\phi = 30°$ and $\phi = 210°$ will produce nonzero amplitudes in both $2\Omega_L$ and $\Omega_L$ resonances for $\theta < 90°$ as shown in Fig. 7(b) (middle & bottom row). The oscillating amplitudes of $2\Omega_L$ and $\Omega_L$ resonances with polarization angle $\phi$ get smaller by tilting the magnetic field $B$ away from the $x$-$y$ plane [Fig. 7(b) (middle and bottom rows)]. Also, for $\theta < 90°$, the oscillation frequency $\Omega_L$ resonance become equal to the frequency of $2\Omega_L$ resonance (i.e. periodicity in $\phi = 180°$). Experimental (dots) and theoretical (solid lines) results in Fig. 7(b) show very good agreement. When angle $\theta = 0°$, the magnetic field $B$ is along the $z$-axis (i.e. longitudinal), which is perpendicular to $E$ for any choice of polarization angle $\phi$. In this case, the amplitude of $2\Omega_L$ resonance does not depend on $\phi$, hence, shows no oscillation. On the other hand, $\Omega_L$ resonance at $\theta = 0°$ completely vanishes due to the absence of $\pi$-transition. The oscillatory behavior showed here in Fig. 7(b) is different from the one observed in Ref. [16] using a second-harmonic i.e. $2\Omega_{mod}$ detection scheme with polarization-modulated light.

Figure 7(c) shows oscillation amplitudes of $2\Omega_L$ and $\Omega_L$ resonances as a function of the magnetic field tilt angle $\theta$. The oscillations show a strong linear dependence over a range of tilt angle $\theta$ from 15° to 40°. Figure 7(c) can be used as a calibration curve for determining angle $\theta$ of the magnetic field, whereas locations of maxima and minima in $2\Omega_L$ and $\Omega_L$ oscillations in Fig. 7(b) can used to find angle $\psi$ of the magnetic field. Compared to vector magnetometer based on a single $\Omega_L$ resonance [43], measurements using both $2\Omega_L$ and $\Omega_L$ resonances can improve accuracy of the vector magnetometer based on the synchronous optical pumping. Similarly, the ratio of relative strengths between $2\Omega_L$ and $\Omega_L$ resonances can be used as a response for avoiding the commonly encountered 'dead-zone' problem in the Bell-Bloom magnetometer [25].

## V. CONCLUSIONS

We have investigated magnetic resonances at nonzero magnetic field created by synchronous optical pumping of the atoms using an OTS coated rubidium vapor cell. The effect of incident light ellipticity on the resonance spectrum is studied in the presence of magnetic field oriented perpendicular to the light propagation direction. Our study showed ground state coherences responsible for producing two types of magnetic resonances, strongly depend on the light ellipticity. Resonance ($\Omega_L = \pm\Omega_{mod}$) satisfying the coherence condition $|\Delta m_g| = 1$ becomes stronger with increase in light ellipticity, whereas resonance ($\Omega_L = \pm\Omega_{mod}/2$) satisfying the coherence condition $|\Delta m_g| = 2$ diminishes at higher ellipticity. We showed that the difference in amplitudes of $\Omega_L = \Omega_{mod}$ and $\Omega_L = \Omega_{mod}/2$ resonances varies linearly with ellipticity between 10º and 30º, which can be used for *in situ* measurement of light ellipticity. For non-zero light ellipticity, we also reported sign reversal in fourth-order dispersive resonance at 30% light duty cycle. We studied the dependence of $2\Omega_L$ and $\Omega_L$ resonance amplitudes on the magnetic field direction using polarization rotation. The amplitudes of $2\Omega_L$ and $\Omega_L$ resonances showed periodic oscillations with polarization rotation. These oscillations are found to be sensitive to the orientation of the magnetic field with respect to plane of polarization. This aspect can be utilized in developing a synchronous optical pumping vector magnetometer.

## ACKNOWLEDGEMENTS

This work is supported by funding received from NASA EPSCoR (80NSSC17M0026), and NASA MIRO (NNX15AP84A).